\documentstyle[12pt]{article}
\def\Tr{\mathop{\rm Tr}\nolimits}
\def\x {\stackrel {\textstyle \otimes}{,}}
\def\beq{\begin{equation}}
\def\eeq{\end{equation}}
\def\be{\begin{displaymath}}
\def\ee{\end{displaymath}}

\catcode`\@=11

\font\teneus=eusm10 scaled \magstep1
\font\seveneus=eusm7 scaled \magstep1
\font\fiveeus=eusm5 scaled \magstep1

\newfam\eusfam
\textfont\eusfam=\teneus  \scriptfont\eusfam=\seveneus
  \scriptscriptfont\eusfam=\fiveeus

\def\hexnumber@#1{\ifnum#1<10 \number#1\else
 \ifnum#1=10 A\else\ifnum#1=11 B\else\ifnum#1=12 C\else
 \ifnum#1=13 D\else\ifnum#1=14 E\else\ifnum#1=15 F\fi\fi\fi\fi\fi\fi\fi}

\def\Cl{\ifmmode\let\next\Cl@\else
 \def\next{\errmessage{Use \string\Cl\space only in math mode}}\fi\next}
\def\Cl@#1{{\Cl@@{#1}}}
\def\Cl@@#1{\fam\eusfam#1}

\catcode`\@=\active

\catcode`\@=11

\font\teneuf=eufm10 scaled \magstep1
\font\seveneuf=eufm7 scaled \magstep1
\font\fiveeuf=eufm5 scaled \magstep1

\newfam\euffam
\textfont\euffam=\teneuf  \scriptfont\euffam=\seveneuf
  \scriptscriptfont\euffam=\fiveeuf

\def\hexnumber@#1{\ifnum#1<10 \number#1\else
 \ifnum#1=10 A\else\ifnum#1=11 B\else\ifnum#1=12 C\else
 \ifnum#1=13 D\else\ifnum#1=14 E\else\ifnum#1=15 F\fi\fi\fi\fi\fi\fi\fi}

\def\Got{\ifmmode\let\next\Got@\else
 \def\next{\errmessage{Use \string\Got\space only in math mode}}\fi\next}
\def\Got@#1{{\Got@@{#1}}}
\def\Got@@#1{\fam\euffam#1}

\catcode`\@=\active

\catcode`\@=11

\font\tenmsx=msxm10 scaled \magstep1
\font\sevenmsx=msxm7 scaled \magstep1
\font\fivemsx=msxm5 scaled \magstep1
\font\tenmsy=msym10 scaled \magstep1
\font\sevenmsy=msym7 scaled \magstep1
\font\fivemsy=msym5 scaled \magstep1
\newfam\msxfam
\newfam\msyfam
\textfont\msxfam=\tenmsx  \scriptfont\msxfam=\sevenmsx
  \scriptscriptfont\msxfam=\fivemsx
\textfont\msyfam=\tenmsy  \scriptfont\msyfam=\sevenmsy
  \scriptscriptfont\msyfam=\fivemsy

\def\hexnumber@#1{\ifnum#1<10 \number#1\else
 \ifnum#1=10 A\else\ifnum#1=11 B\else\ifnum#1=12 C\else
 \ifnum#1=13 D\else\ifnum#1=14 E\else\ifnum#1=15 F\fi\fi\fi\fi\fi\fi\fi}

\def\Bbb{\ifmmode\let\next\Bbb@\else
 \def\next{\errmessage{Use \string\Bbb\space only in math mode}}\fi\next}
\def\Bbb@#1{{\Bbb@@{#1}}}
\def\Bbb@@#1{\fam\msyfam#1}

\catcode`\@=\active

\newtheorem{theorem}{Theorem}[section]
\newtheorem{lemma}[theorem]{Lemma}

\newtheorem{definition}[theorem]{Definition}

\newenvironment{proof}{\begin{trivlist} \item[] {\em Proof}. }%
{\hfill $\Box$ \end{trivlist}}

\newcounter{exercise}

{\end{trivlist}\normalsize}

\title{
$R$-matrices for Elliptic Calogero-Moser Models
}
\author{H. W. Braden
\quad\quad Takashi Suzuki\\
\\
\normalsize
Department of Mathematics and Statistics,\\
\normalsize
University of Edinburgh,\\
\normalsize
Edinburgh, UK.
}

\begin{document}
\begin{titlepage}
\renewcommand{\thepage}{}
\maketitle
\vskip-9.5cm
\hskip11.4cm
Edinburgh-/92-93/03
\vskip.2cm
\hskip11.4cm
\sf hep-th/9309033 \rm
\vskip8.8cm

\begin{abstract}
The classical $R$-matrix structure for the $n$-particle Calogero-Moser
models with (type IV) elliptic potentials is investigated. We show
there is no momentum independent $R$-matrix (without spectral parameter)
when $n\ge4$. The assumption of momentum independence is sufficient
to reproduce the dynamical $R$-matrices of Avan and Talon for the
type I,II,III degenerations of the elliptic potential.
The inclusion of a spectral parameter enables us to find
$R$-matrices for the general elliptic potential.
\end{abstract}
\vfill
\end{titlepage}
\renewcommand{\thepage}{\arabic{page}}

\section{Introduction}

The Calogero-Moser model \cite{Cal75,Moser} is perhaps the
paradigm of a completely
integrable system of $n$-particles on the line  which interact via
pairwise potentials $v(q_i-q_j)$.
The most general form of the potential in such models \cite{OlP81} is the
(so called type IV) elliptic potential, $v(q)={a\sp2}\wp(a q)$, where
$\wp$ is the Weierstrass elliptic function.
The various degenerations of this function yield
rational (type I, $v(q)=1/{q\sp2}$),
hyperbolic (type II, $v(q)=a\sp2/{\sinh(a q)\sp2}$) and
trigonometric (type III, $v(q)=a\sp2/{\sin(a q)\sp2}$) potentials.
In accord with their importance these models have been studied from
many different perspectives. They have a Lax pair formulation,
$i {\dot L}= [L,M]$; the ansatz for this Lax pair leads to a study of
functional equations\cite{Cal76,Ruijs87,BC90,BB93}.
Further, these models may be expressed as the Hamiltonian reduction
\cite{KKS,ABT} of
integrable flows on the cotangent bundle of symmetric spaces.
The quantisation \cite{OlP83} of these models has also been of some interest.
Recently the related (type V) potential $v(q)=1/{q\sp2}+g{q\sp2}$ has
been shown  to be relevant to the collective field theory of
strings\cite{AJ91}.

The proof of the complete integrability of a system given in terms of a Lax
pair involves several stages. The first, an immediate consequence of a
Lax pair formulation, is the observation that the quantities
$\Tr_E {L\sp k}$ are conserved. Here the trace is taken over the representation
$E$ of the Lie algebra ${\Got{g}}$ to which the operator $L$ is
associated. Another stage is to show these provide enough functionally
independent conserved quantities.
Finally, and this is perhaps the most tedious step,
one  must show the quantities are in involution, {\it i.e.}
$\{\Tr_E {L\sp k},\Tr_E {L\sp m}\}=0$.
This step is model dependent. For the Calogero-Moser this stage may
be achieved by arguments based on asymptotics\cite{Moser},
inverse scattering\cite{OlP76} or direct recursion\cite{W77}.
Given an $L$-operator, an alternative approach to proving this Poisson
commutativity proceeds via the $R$-matrix\cite{STS,FT,BV90}.
An $R$-matrix is an $E\otimes E$ matrix satisfying
\beq
\{L\x L\}=[R, L\otimes 1]-[{R\sp\pi},1\otimes L] .
\label{rmatrix}
\eeq
An immediate consequence of the existence of an $R$-matrix is that
$$\{\Tr_E {L\sp k},\Tr_E {L\sp m}\}=
  \Tr_{E\otimes E}\{ {L\sp k}\x {L\sp m}\}=
  k m \Tr_{E\otimes E} {L\sp {k-1}}\otimes{L\sp {m-1}} \{L\x L\}=
% k m \Tr_{E\otimes E} {L\sp {k-1}}\otimes{L\sp {m-1}} (
% [R, L\otimes 1]-[{R\sp\pi},1\otimes L] )=
  0.
$$
The vanishing
follows from (\ref{rmatrix}) by expressing $\{ L \x L\}$ as a commutator
and using the cyclicity of the trace. An $R$-matrix also allows one to
canonically construct the matrix $M$ of the Lax pair\cite{STS}.
Recently Avan and Talon\cite{AT}
constructed the $R$-matrices for the Calogero-Moser
models and potentials of type I, II, III and V, thus providing this alternate
means of proof.
The $R$-matrices found by Avan and Talon were dynamical:
that is, they depended on the dynamical variables (in their case positions)
of the model. In contrast to systems governed by purely numerical
$R$-matrices\cite{FT}, dynamical $R$-matrices needn't satisfy the Yang-Baxter
equation and the theory of dynamical $R$-matrices is not well
 understood\cite{Maillet,EEKT,BB}.
One hopes that such concrete examples of dynamical $R$-matrices as are provided
by the Calogero-Moser models will aid in the elucidation of this theory.

Our present work investigates the $R$-matrix  structure
of the Calogero-Moser models further.
The $R$-matrices of Avan and Talon were constructed on the basis of two
assumptions, namely  momentum independence and the vanishing of certain terms
of the $R$-matrix.
These assumptions were found to be consistent with the
potentials of type I, II, III, V but did not allow the type IV potential.
One might ask what happens if these assumptions are relaxed.
We shall show  that the second of Avan and Talon's assumptions actually
follows from that of momentum independence (given sufficient
particles). We can  therefore conclude that  no momentum independent
$R$-matrix exists for the $L$-operators under consideration.
In \cite{Krich} Krichever enlarged the class of $L$-operators
yielding type IV potentials to include a spectral parameter.
The usual $L$-operators exactly correspond to those values of the
spectral parameter for which the operators are hermitian.
We show that the inclusion of the spectral parameter allows us
to construct a momentum independent $R$-matrix.

The plan of this paper is as follows. In Section 2  we will introduce
our notation, the $L$-operators under consideration and the
equations to be solved. Section 3 looks at the simplifications
resulting from the assumption of momentum independence.
Section 4 then shows that for the usual $L$-operator no
momentum independent $R$-matrix can be found.
Upon introducing a spectral parameter in Section 5,
we then
exhibit a solution to the corresponding equations.
 We conclude with a brief discussion.

\section{Preliminaries}
As an alternative to the matrix entry calculations often presented,
we give our calculations in terms of a basis of the underlying
Lie algebra.
Although we will ultimately specialise to the $gl_n$ case, we
believe this to be both computationally and
conceptually easier. It will also enable us to isolate
those features peculiar to  $gl_n$.
We begin by deriving
the  relevant equations to be solved and introducing
our notation.

Let $X_\mu$ denote a Cartan-Weyl basis for the
(semi-simple) Lie algebra ${\Got{g}}$
associated with the operator $L$. That is $\{X_\mu\}=\{H_i,E_\alpha\}$,
where $\{H_i\}$ is a basis for the Cartan subalgebra ${\Got{h}}$
and $\{E_\alpha\}$ is the set of  step operators (labelled by
the root system $\Phi$ of ${\Got{g}}$) and
$$
[H_i,E_\alpha ]=\alpha_i E_\alpha \quad\quad
[E_\alpha,E_{-\alpha}]= \alpha\cdot H\quad\quad
{\rm and}\quad\quad [E_\alpha,E_\beta ]=N_{\alpha,\beta}E_{\alpha+\beta}
\quad{\rm if}\ \alpha+\beta\in\Phi .
$$
With $[X_\mu,X_\nu]=c_{\mu \nu}\sp\lambda\ X_\lambda$ defining the
structure constants of ${\Got{g}}$, we see  for example\footnote{
Throughout, Roman indices will denote the Cartan subalgebra basis elements
while the early Greek indices $\alpha,\beta,\ldots$
will denote the  step operators.}
that
$c_{i\,\alpha}\sp\beta=\delta_{\alpha,\beta}\alpha_i$
and $c_{\alpha\, -\alpha}\sp i =\alpha_i$.
Further, we adopt the hermiticity convention ${E_\alpha}\sp\dagger=E_{-\alpha}$
for the representations of ${\Got{g}}$. The structure constants may then
be chosen to be real and  to have the symmetries:
$c_{\alpha\, \beta}\sp\lambda = -c_{\beta\,\alpha}\sp\lambda =
c_{\lambda\, -\beta}\sp\alpha= -c_{-\alpha\, -\beta}\sp{-\lambda}$.

With this notation at hand we express the $L$ operator as
\be
L\equiv \sum_{\mu}L\sp\mu X_\mu = p\cdot H + i\sum_{\alpha\in \Phi}w_\alpha
E_\alpha
\ee
where\cite{Krich}
\beq
w_\alpha\equiv w(\alpha\cdot q;u)=
{\sigma(u-\alpha\cdot q)\over{\sigma(u)\sigma(\alpha\cdot q)}}
{e\sp{\zeta(u)\alpha\cdot q}}.
\label{eq:w}
\eeq
Here $\sigma(x)$ and $\zeta(x)=\sigma\sp\prime(x)/ \sigma(x)$
are the Weierstrass sigma and zeta functions\cite{WW}.
The quantity $u$ in (\ref{eq:w}) is known as the spectral
parameter and we will only make its appearance explicit when confusion
might otherwise arise.
It will also be convenient to use the shorthand $f_\alpha$ for a
function on ${\Got{h}}$ that takes the value
$f(\alpha\cdot q)$  when evaluated at $q$.
The functions $w_\alpha$ satisfy the addition formula
\beq
w_\alpha w_\beta\sp\prime -w_\beta w_\alpha\sp\prime=
( z_\alpha-z_\beta)w_{\alpha+\beta},
\quad\quad{\rm where}\quad
 z_\alpha(u)={w_\alpha\sp{\prime\prime}(u)\over {2 w_\alpha(u)}}=
\wp_\alpha +{1\over2}\wp(u) .
\label{eq:addition}
\eeq
Clearly $L=L\sp\dagger$ when $w_\alpha=-w_{-\alpha}$ and this  is
the case usually
considered.
Requiring hermiticity restricts the spectral parameter with the
result  that
$u\in\{\omega_1,\omega_2,-\omega_1-\omega_2\}$, where $2\omega_{1,2}$ are
the periods of the associated elliptic functions.

For the Lie algebra $gl_n$, the case of most interest to us,
$\Phi=\{e_i-e_j,\ 1\leq i\not= j\leq n\}$ with the $e_i$  an orthonormal
basis of ${\Bbb R}\sp n$.
If $e_{rs}$ denotes the elementary matrix with $(r,s)-th$ entry one and
zero elsewhere, then the $n\times n$ matrix representation $H_i=e_{ii}$
and $E_\alpha=e_{ij}$ when $\alpha=e_i-e_j$ gives the usual
representation of $L$. Working with the simple algebra $a_n$ corresponds
to the center of mass frame.

Let us begin unravelling (\ref{rmatrix}). The
left hand side becomes
\beq
\{L\x L\}=\sum_{\mu,\nu} \{ L\sp\mu, L\sp\nu \} X_\mu\otimes X_\nu
=i\sum_{j,\alpha}\alpha_j w_\alpha\sp\prime (H_j\otimes E_\alpha-
E_\alpha\otimes H_j)
\label{rmrhs}
\eeq
upon using
$\{p_j,w_\alpha\}=
 \{p_j,\alpha\cdot q\}w_\alpha\sp\prime=\alpha_j w_\alpha\sp\prime$.
Turning now  to the right hand side of (\ref{rmatrix}) we have
$R=R\sp{\mu\nu}X_\mu\otimes X_\nu$ and
${R\sp\pi}=R\sp{\nu\mu}X_\mu\otimes X_\nu$. Then
\begin{eqnarray}
[R, L\otimes 1]-[{R\sp\pi},1\otimes L]&=&
R\sp{\mu\nu}([X_\mu,L]\otimes X_\nu-X_\nu\otimes[X_\mu,L])\cr
&=& R\sp{\mu\nu}L\sp\lambda
 ([X_\mu,X_\lambda]\otimes X_\nu-X_\nu\otimes[X_\mu,X_\lambda])\cr
&=&( R\sp{\tau\nu}c_{\tau\lambda}\sp\mu L\sp\lambda -
    R\sp{\tau\mu}c_{\tau\lambda}\sp\nu L\sp\lambda) X_\mu\otimes X_\nu.
\nonumber
\end{eqnarray}
In terms of the Lie algebra basis, (\ref{rmatrix}) then becomes the
equation
\beq
\{ L\sp\mu, L\sp\nu \}= R\sp{\tau\nu}c_{\tau\lambda}\sp\mu L\sp\lambda -
  R\sp{\tau\mu}c_{\tau\lambda}\sp\nu L\sp\lambda .
\label{rmatrixnew}
\eeq
Observe  that (\ref{rmatrixnew}) has the structure of a matrix equation,
\be
V R +R\sp\dagger V =A,
\ee
for the unknown matrix $R$ in terms of the specified
$A\sp{\mu\nu}=\{ L\sp\mu, L\sp\nu \}$ and
$V\sp{\mu\nu}= c_{\nu\lambda}\sp\mu L\sp\lambda$.
Equation (\ref{rmatrixnew}) yields three different equations,
depending on the range of $\{\mu,\nu\}$. For
$(\mu,\nu)= (i,j),(i,\alpha)$ and $(\alpha,\beta)$ respectively, these are
\begin{eqnarray}
\label{eq:ij}
0&=& \sum_{\alpha}(R\sp{\alpha j}\alpha_i-R\sp{\alpha i}\alpha_j)w_{-\alpha}\\
\label{eq:ia}
-\alpha_i w_\alpha\sp\prime &=&
i\alpha\cdot p\ R\sp{\alpha i} +\alpha\cdot R\sp{i}w_\alpha +
      \sum_{\beta}(\beta_i w_\beta R\sp{-\beta\alpha } +
       w_{\alpha -\beta} R\sp{\beta i}c_{\beta\ \alpha -\beta}\sp\alpha)  \\
\noalign{\hbox{and}}\cr
\label{eq:ab}
0&=& \alpha\cdot R\sp{\beta}w_\alpha -\beta\cdot R\sp{\alpha}w_\beta
  +i ( \alpha\cdot p\ R\sp{\alpha\beta}-\beta\cdot p\ R\sp{\beta\alpha})\\
\nonumber
&&  +\sum_{\gamma}(
   R\sp{\gamma\beta} c_{\gamma\, \alpha-\gamma}\sp\alpha w_{\alpha-\gamma}
  -R\sp{\gamma\alpha} c_{\gamma\, \beta-\gamma}\sp\beta w_{\beta-\gamma}
)
\end{eqnarray}
Here we have introduced the shorthand
$\beta\cdot R\sp{\mu}\equiv\sum_{i}\beta_i R\sp{i\mu}$.
Equations (\ref{eq:ij}-\ref{eq:ab}) are the components of
(\ref{rmatrix}) in our basis.

\section{Momentum Independence}

We now turn to the solution of equations (\ref{eq:ij}-\ref{eq:ab})
subject to  { the assumption that the $R$-matrix is independent
of momentum}. This assumption (introduced in \cite{AT}) means that
\beq
R\sp{\alpha i}=0,\quad\quad
R\sp{\alpha\, -\alpha}+R\sp{-\alpha\, \alpha}=0\quad{\rm and}\quad
R\sp{\alpha\beta}=0\quad{\rm if}\quad\alpha\not=\pm\beta.
\label{eq:momcons}
\eeq
The first of these  restrictions follows from (\ref{eq:ia}) while the
remainder come from (\ref{eq:ab}). For example, in the matrix components
of $gl_n$ introduced earlier, we have
$R\sp{\alpha i}=0\Leftrightarrow R\sp{jkii}=0$ and we thus obtain equations
(14)
of \cite{AT}. At this stage,  equation (\ref{eq:ij}) is  satisfied identically
and the variables remaining are
$R\sp{ij},R\sp{i\alpha},R\sp{\alpha\alpha}$ and $R\sp{\alpha\, -\alpha}$.
The remaining equations to be solved are
\begin{eqnarray}
-\alpha_i w_\alpha\sp\prime &=&
\alpha\cdot R\sp{i}w_\alpha + \alpha_i R\sp{-\alpha\alpha}w_\alpha
 -\alpha_i R\sp{\alpha\alpha}w_{-\alpha}
\label{eq:ian}
\\
\noalign{\hbox{and}}\cr
\alpha\cdot R\sp{\beta}w_\alpha -\beta\cdot R\sp{\alpha}w_\beta &=&
c_{\alpha\gamma}\sp\beta(
  R\sp{\alpha\alpha}w_\gamma-R\sp{\beta\beta}w_{-\gamma}) +
c_{-\alpha\gamma}\sp\beta(
  R\sp{-\alpha\alpha}w_\gamma+R\sp{-\beta\beta}w_{\gamma}).
\label{eq:abn}
\end{eqnarray}
The first term on the right-hand side of (\ref{eq:abn}) is nonvanishing
only for $\gamma=\beta-\alpha\in\Phi$ while the second term is  nonvanishing
for
$\gamma=\beta+\alpha$.
We note that for the simply-laced algebras
($\alpha\cdot\alpha=2,\ \forall\ \alpha\in\Phi$),
at most one of the terms on the
right-hand of (\ref{eq:abn}) can  be nonvanishing and we
henceforth assume this to be the case.
Now, by viewing the root $\gamma=\beta-\alpha$ as being also the
sum  $\gamma=\beta+(-\alpha)$, we obtain the two
equations
\begin{eqnarray}
\label{eq:g-}
\alpha\cdot R\sp{\beta}w_\alpha -\beta\cdot R\sp{\alpha}w_\beta &=&
c_{\alpha\gamma}\sp\beta(
  R\sp{\alpha\alpha}w_\gamma-R\sp{\beta\beta}w_{-\gamma}) \\
\noalign{\hbox{and}}\cr
\label{eq:g+}
-\alpha\cdot R\sp{\beta}w_{-\alpha} -\beta\cdot R\sp{-\alpha}w_\beta &=&
c_{\alpha\gamma}\sp\beta(
R\sp{\alpha-\alpha}w_\gamma+R\sp{-\beta\beta}w_{\gamma}).
\end{eqnarray}
We shall utilise the consistency of these equations below.

Our first observation is that $R\sp{ij}=\eta\,\delta\sp{ij}+P\sp{ij}$ for
some constant $\eta$ that we shall later determine and matrix $P\sp{ij}$
orthogonal to the roots, $\alpha\cdot P\sp{j}=0\,\forall\,j$.
To see this, view (\ref{eq:ian}) as an equation between vectors;
thus $\alpha\cdot R\sp{i}$ must be proportional to $\alpha_i$. If
we define the constant of proportionality by
$\alpha\cdot R\sp{i}=\eta_\alpha\alpha_i$, where $\eta_\alpha$ could
in principle depend on $\alpha$, then by linearity
$$(\alpha+\beta)\cdot R\sp{i}=
\eta_{\alpha+\beta}(\alpha_i+\beta_i)=\eta_\alpha\alpha_i+\eta_\beta\beta_i
$$
and so $\eta_{\alpha+\beta}=\eta_\alpha=\eta_\beta\equiv\eta$.
Now for each $\alpha$  we have $\sum_i \alpha_iR\sp{ij}=\eta\alpha_j$,
and so $R\sp{ij}=\eta\,\delta\sp{ij}+P\sp{ij}$. The matrix $P\sp{ij}$
orthogonal to the roots arises when we have $u(1)$ factors present in
${\Got{g}}$.
Thus we have
\beq
-w_\alpha\sp\prime=\eta w_\alpha + R\sp{-\alpha\alpha}w_\alpha
 - R\sp{\alpha\alpha}w_{-\alpha}.
\label{eq:iann}
\eeq
The assumption of momentum independence leads then to the two equations
(\ref{eq:abn}) and (\ref{eq:iann}).

\section{The case $w_\alpha=-w_{-\alpha}$.}

Our discussion has so far made no use of the form of $w_\alpha$.
For the remainder of this section we will assume that $w_\alpha$ is an
odd function, the case usually considered. As we shall see, this
results in some quite strong conclusions. The next section,  which deals
with the inclusion of a spectral parameter, will consider the more general
case.
First let us show
\begin{lemma} $\eta=0$.
\end{lemma}
\begin{proof}
Upon subtracting from (\ref{eq:iann}) the analogous equation  obtained
by replacing $\alpha$ with $-\alpha$
and using the second equation of (\ref{eq:momcons}), we find
\beq
\eta=-{1\over2}(R\sp{\alpha\alpha}+R\sp{-\alpha-\alpha}).
\label{eq:eta}
\eeq
Further, upon subtracting (\ref{eq:g-}) from (\ref{eq:g+}), we obtain
\begin{eqnarray}
\beta\cdot(R\sp{\alpha}-R\sp{-\alpha})w_\beta &=&
c_{\alpha\gamma}\sp\beta(
R\sp{\alpha-\alpha}+R\sp{-\beta\beta}-R\sp{\alpha\alpha}-R\sp{\beta\beta})
w_\gamma .\\
\noalign{\hbox{
The same operations applied to the analogous equations based now on
$\gamma=(-\alpha)-(-\beta)$}}
\noalign{\hbox{ and $-\gamma=\alpha-\beta$ yield }}\cr
\beta\cdot(R\sp{\alpha}-R\sp{-\alpha})w_\beta &=&
c_{\alpha\gamma}\sp\beta(
R\sp{-\alpha-\alpha}+R\sp{-\beta-\beta}-R\sp{-\alpha\alpha}-R\sp{\beta-\beta})
w_\gamma .
\end{eqnarray}
Upon comparing these last two equations and using (\ref{eq:eta}) together
with
$R\sp{\alpha\, -\alpha}+R\sp{-\alpha\, \alpha}=0$, we find that $\eta=0$.
\end{proof}

Therefore $R\sp{ij}=P\sp{ij}$; the choice
$P\sp{ij}=0$ corresponds to the middle two equations of Avan and
Talon's second assumption\cite{AT}.
We have now reduced the possible nonzero variables of the $R$-matrix to
$R\sp{i\alpha},R\sp{\alpha\alpha}$ and $R\sp{\alpha\, -\alpha}$ subject
to
\beq
R\sp{\alpha\, -\alpha}+R\sp{-\alpha\, \alpha}=0,\quad\quad
R\sp{\alpha\alpha}+R\sp{-\alpha\,-\alpha}=0,\quad\quad
R\sp{\alpha\alpha}+R\sp{-\alpha\, \alpha}=-
{w_\alpha\sp\prime \over w_\alpha}
\label{eq:redcons}
\eeq
and
\beq
\alpha\cdot(R\sp{\beta}-R\sp{-\beta})w_\alpha=c_{\alpha\gamma}\sp\beta
\big(
-{w_\alpha\sp\prime \over w_\alpha} -{w_\beta\sp\prime \over w_\beta}
+2 R\sp{\alpha\, -\alpha}
\big)w_\gamma
-c_{-\alpha\gamma\sp\prime}\sp\beta\big(
-{w_\alpha\sp\prime \over w_\alpha} +{w_\beta\sp\prime \over w_\beta}
+2 R\sp{\alpha\, -\alpha}\big)w_{\gamma\sp\prime}.
\label{eq:add}
\eeq
Of course $c_{\alpha\gamma}\sp\beta=0$ unless $\gamma=\beta-\alpha\in\Phi$
(in which case $\alpha\cdot\gamma\not=0$).
Our method of solving these equations proceeds as follows.
Let us define the  quantity $A_{\beta\gamma}$ by
\begin{definition}
\beq
A_{\beta\gamma}\equiv
\beta\cdot(R\sp{\gamma}-R\sp{-\gamma})w_\gamma
-c_{\gamma-\beta\,\beta}\sp\gamma z_{\gamma}.
\eeq
\end{definition}
Obviously this is closely related
to the left hand side of (\ref{eq:add}).
Our aim will be to show this
quantity  to be constant, from which we will be able to deduce
the remaining equations of Avan and Talon's second assumption.

\begin{lemma}
$
A_{\alpha\beta}=A_{\beta\alpha}=-A_{\beta\gamma}=A_{\gamma-\alpha}=
A_{-\beta\gamma}.
$
\end{lemma}
\begin{proof}
First let us motivate the definition of $A_{\beta\gamma}$ and then
derive its symmetries.
Suppose $\gamma=\beta-\alpha\in\Phi$ so only the first term of
(\ref{eq:add}) is nonvanishing.
The analogous equation  to (\ref{eq:add}) for $\beta=(\beta-\alpha)-(-\alpha)$
is
\beq
\alpha\cdot(R\sp{\gamma}-R\sp{-\gamma})w_\alpha=
c_{-\alpha\beta}\sp\gamma\big(
{w_\alpha\sp\prime \over w_\alpha} -{w_\gamma\sp\prime \over w_\gamma}
+2 R\sp{-\alpha\alpha}
\big)w_\beta .
\label{eq:addp}
\eeq
Upon adding $w_\beta \times(\ref{eq:add})$ to $w_\gamma\times(\ref{eq:addp})$
and using the symmetries of the structure constants
together with the addition formula (\ref{eq:addition}) for $w_\alpha$, we
obtain
$$
\alpha\cdot(R\sp{\beta}-R\sp{-\beta})w_\beta+
\alpha\cdot(R\sp{\gamma}-R\sp{-\gamma})w_\gamma=
c_{\alpha\gamma}\sp\beta (z_{\gamma}-z_\beta).
$$
(For the case at hand, $z_{\gamma}=z_{-\gamma}$.)
After substituting $\alpha=\beta-\gamma$ in this expression and
making use of the fact $\alpha\cdot(R\sp{\alpha}-R\sp{-\alpha})=0$, which
follows from (\ref{eq:add}), we  obtain
\beq
\beta\cdot(R\sp{\gamma}-R\sp{-\gamma})w_\gamma
-c_{-\alpha\beta}\sp\gamma z_{\gamma}=
\gamma\cdot(R\sp{\beta}-R\sp{-\beta})w_\beta
-c_{\alpha\gamma}\sp\beta z_\beta .
\label{eq:aas}
\eeq
Therefore $A_{\beta\gamma}=A_{\gamma\beta}$.
Upon using $\alpha=\beta-\gamma$ the remaining symmetries are
similarly shown.
\end{proof}

Thus to every triangle formed by three roots  $\alpha,\beta,\gamma$
we have associated a single
constant  $A_{\alpha\beta}$
(up to a sign which is taken care of below).

Now suppose the root $\alpha$ may be expressed as a sum of two distinct
pairs of roots, $\alpha=\beta-\gamma=\beta\sp\prime-\gamma\sp\prime$.
This requires  that $n\ge4$.
For the simply-laced case being considered, we may further
assume $\alpha\cdot(\gamma-\gamma\sp\prime)=0$
and that our labelling is such that $\gamma-\gamma\sp\prime\in\Phi$.
What then is the relation between $A_{\alpha-\gamma}$ and
$A_{\alpha{-\gamma\sp\prime}}$?
Using (\ref{eq:add}) we see that
\beq
(\gamma-\gamma\sp\prime)\cdot(R\sp{\alpha}-R\sp{-\alpha})=0.
\label{eq:cons}
\eeq
Further
$c_{\beta -\gamma}\sp{\alpha}=
 c_{\beta\sp\prime -\gamma\sp\prime}\sp{\alpha}$,
and so
\begin{displaymath}
A_{\alpha-\gamma}=-\gamma\cdot(R\sp{\alpha}-R\sp{-\alpha})w_\alpha
-c_{\beta -\gamma}\sp{\alpha}z_\alpha  =
-\gamma\sp\prime\cdot(R\sp{\alpha}-R\sp{-\alpha})w_\alpha
-c_{\beta\sp\prime -\gamma\sp\prime}\sp{\alpha}z_\alpha=
A_{\alpha-\gamma\sp\prime} .
\end{displaymath}
We have just shown that the constants $A_{\alpha-\gamma}$ associated with
a triangle of roots are the same whenever they share a root $\alpha$ as
a common edge. Now we can get from one triangle of roots to
any other by intermediate root triangles. Therefore the constants
$A_{\alpha-\gamma}$ depend on all the roots in the same fashion
and we  have shown
\begin{lemma}
$A_{\alpha\beta}=c_{\beta-\alpha\,\alpha}\sp{\beta}{\cal A} $\,
for some function ${\cal A}$.
\end{lemma}

On combining this lemma and (\ref{eq:add}) we see that
\beq
\alpha\cdot(R\sp{\beta}-R\sp{-\beta})w_\alpha w_\beta
=c_{\gamma\alpha}\sp\beta
\big(
{w_\alpha\sp\prime \over w_\alpha} +{w_\beta\sp\prime \over w_\beta}
+2 R\sp{-\alpha \alpha}
\big)w_\gamma w_\beta =
c_{\gamma\alpha}\sp{\beta}({\cal A}+z_\beta)w_\alpha .
\label{eq:sol}
\eeq
Thus  ${\cal A}$ determines $R\sp{-\alpha \alpha}$ and
(via (\ref{eq:redcons})) $R\sp{\alpha \alpha}$,
assuming we are given $w_\alpha$ and $z_\alpha$. Further,
\begin{lemma} For $n\ge4$ ${\cal A}$ is a constant and $ R\sp{-\alpha \alpha}$
is a function of $\alpha$ only.
\end{lemma}
\begin{proof}
Once again, suppose the root $\alpha$ may be expressed as a sum of two distinct
pairs of roots, $\alpha=\beta-\gamma=\beta\sp\prime-\gamma\sp\prime$.
Comparing (\ref{eq:sol})  with the analogous equation in $\beta\sp\prime$,
$\gamma\sp\prime$ enables us to show that
\begin{displaymath}
\big(
{w_\beta\sp\prime \over w_\beta} -{{z_\beta w_\alpha}\over{w_\gamma w_\beta}}
 \big) - \big(
{w_{\beta\sp\prime}\sp\prime \over w_{\beta\sp\prime}}-
{{z_{\beta\sp\prime}w_\alpha}\over{w_{\gamma\sp\prime}w_{\beta\sp\prime}} }
\big) = \big(
{w_\alpha\over{w_\gamma w_\beta}}-
{w_\alpha\over{w_{\gamma\sp\prime}w_{\beta\sp\prime}}}
\big){\cal A} ,
\end{displaymath}
and so we may solve for ${\cal A}$ explicitly in terms of the roots shown.
We have argued however that ${\cal A}$ depends on all of the roots in the
same fashion. Therefore ${\cal A}$ is a constant.
Having shown  that ${\cal A}$ is a constant, let us rewrite (\ref{eq:sol})
again, assuming that
$\alpha=\beta-\gamma=\beta\sp\prime-\gamma\sp\prime$. Then,
\begin{displaymath}
({\cal A}+z_\beta){{w_\alpha}\over{w_\gamma w_\beta}}-
{w_\beta\sp\prime \over w_\beta} =
2 R\sp{-\alpha \alpha}+{w_\alpha\sp\prime \over w_\alpha}=
({\cal A}+z_{\beta\sp\prime})
{{w_\alpha}\over{w_{\gamma\sp\prime}w_{\beta\sp\prime}}}-
{w_{\beta\sp\prime}\sp\prime \over w_{\beta\sp\prime}}.
\end{displaymath}
The right-hand side of this equation is a function of $\beta$ and $\gamma$
only, while the left is a function of $\beta\sp\prime$ and
$\gamma\sp\prime$. Thus both are functions of $\alpha$ only and
the remainder of the lemma follows.
\end{proof}

The final stage of our argument consists of showing there is no constant
${\cal A}$ which makes
$2 R\sp{-\alpha \alpha}+{w_\alpha\sp\prime / w_\alpha}$
a function of $\alpha$ only for the elliptic potentials being considered.
Take for example\cite{WW} $w_\alpha=1/sn\,(\alpha\cdot x,k)$. Here
\begin{displaymath}
({\cal A}+z_\beta){{w_\alpha}\over{w_\gamma w_\beta}}-
{w_\beta\sp\prime \over w_\beta} =
({\cal A}-{{1+k\sp2}\over2}+{k\sp2\over w_\alpha\sp2 })
{{w_\alpha}\over{w_\gamma w_\beta}}-{w_\alpha\sp\prime \over w_\alpha},
\end{displaymath}
and we cannot both have ${\cal A}$ constant and this expression
depending  only on $\alpha$  unless $k=0$.  For $k=0$,
which corresponds to the type III degeneration, we find
$R\sp{-\alpha \alpha}=-{w_\alpha\sp\prime / w_\alpha}$,
$R\sp{\alpha \alpha}=0$ and
\begin{displaymath}
\alpha\cdot(R\sp{\beta}-R\sp{-\beta})
=c_{\gamma\alpha}\sp\beta w_\beta .
\end{displaymath}
For $gl_n$, this has a solution $R\sp{i\beta}=w_\beta/2$ when
$\beta\cdot e_i \not=0$, and zero otherwise. We have thus obtained the
remainder of Avan and Talon's assumptions together with their
solution\cite{AT}.  We have therefore shown
\begin{theorem}  If $n\ge4$ there are no momentum
independent $R$-matrices for the nondegenerate type IV potential
and $w_\alpha=-w_{-\alpha}$.
\end{theorem}
When $n=2,3$ our consistency arguments do not arise. For
$n=2$ there is only one root and $n=3$ only one root triangle
and solutions in both cases are possible.

\section{Inclusion of a Spectral Parameter}
Having shown there are no momentum independent $R$-matrices for
$L$-operators with $w_\alpha=-w_{-\alpha}$, several possibilities remain.
We may  for example relax the assumption of momentum independence and solve the
full equations (\ref{eq:ij}-\ref{eq:ab}), or we may look at a broader
class of functions $w_\alpha$. We will adopt the latter approach in this
note and seek momentum independent $R$-matrices for the
the class of  $L$-operators (introduced by Krichever)
containing a spectral parameter. The generalisation of
(\ref{rmatrix}) to the situation with spectral parameter is
\beq
\{L(u)\x L(v)\}=[R(u,v), L(u)\otimes 1]-[{R\sp\pi}(u,v),1\otimes L(v)].
\label{rmatrixsp}
\eeq
If $R(u,v)=R\sp{\mu\nu}(u,v)X_\mu\otimes X_\nu$ then
${R\sp\pi}(u,v)$  is defined by
${R\sp\pi}(u,v)=R\sp{\nu\mu}(v,u)X_\mu\otimes X_\nu$.

We proceed in the same manner given earlier. The left-hand side of
(\ref{rmatrixsp}) is
\be
\{L(u)\x L(v)\}=
i\sum_{j,\alpha}\Bigl(\alpha_j w_\alpha\sp\prime(v) H_j\otimes E_\alpha-
\alpha_j w_\alpha\sp\prime(u) E_\alpha\otimes H_j\Bigl),
\ee
and in terms of our basis (\ref{rmatrixsp}) takes the form
\beq
\{ L\sp\mu(u), L\sp\nu(v) \}=
R\sp{\tau\nu}(u,v)c_{\tau\lambda}\sp\mu L\sp\lambda(u) -
R\sp{\tau\mu}(v,u)c_{\tau\lambda}\sp\nu L\sp\lambda(v) .
\eeq
Again three equations arise, depending on the range of $\{\mu,\nu\}$.
(The new possibility $(\mu,\nu)=(\alpha,i)$ yields the same
equation as $(\mu,\nu)=(i,\alpha)$ with $u$ and $v$ interchanged.)

Once again the assumption that $R(u,v)$ is momentum independent
greatly  reduces the possible nonzero components of $R(u,v)$. We find
\beq
R\sp{\alpha i}(u,v)=0,\quad\quad
R\sp{\alpha\, -\alpha}(u,v)+R\sp{-\alpha\, \alpha}(v,u)=0\quad{\rm and}\quad
R\sp{\alpha\beta}(u,v)=0\quad{\rm if}\quad\alpha\not=\pm\beta.
\eeq
The components to be determined are
$R\sp{ij}(u,v),R\sp{i\alpha}(u,v),R\sp{\alpha\alpha}(u,v)$ and
$R\sp{\alpha -\alpha}(u,v)$.
Again we may argue  that $R\sp{ij}(u,v)=\eta(u,v)\,\delta\sp{ij}+P\sp{ij}$,
and we arrive
at two equations
\beq
\label{eq:spa}
-w_\alpha\sp\prime(v)=\eta(u,v)
 w_\alpha(v) + R\sp{-\alpha\alpha}(u,v)w_\alpha(u)
 - R\sp{\alpha\alpha}(v,u)w_{-\alpha}(u)
\eeq
and
\begin{eqnarray}
\label{eq:spb}
\alpha\cdot R\sp{\beta}(u,v)w_\alpha(u) -\beta\cdot R\sp{\alpha}(v,u)
w_\beta(v)&=&
c_{\alpha\gamma}\sp\beta\Bigl(
  R\sp{\alpha\alpha}(v,u)w_\gamma(v)-R\sp{\beta\beta}(u,v)w_{-\gamma}(u)
\Bigl)\phantom{1234} \\
\nonumber
&&+
c_{-\alpha\gamma}\sp\beta\Bigl(
  R\sp{-\alpha\alpha}(v,u)w_\gamma(v)+R\sp{-\beta\beta}(u,v)w_{\gamma}(u)
\Bigl).
\end{eqnarray}
These equations are the analogues of (\ref{eq:iann}) and (\ref{eq:abn})
respectively.

At this stage we make the ansatz
\beq
\eta(u,v)=\zeta(v-u)+\zeta(u)-\zeta(v),\quad\quad
R\sp{\alpha\alpha}(v,u)=0\quad\quad
%{\rm and}\quad\quad
\label{eq:ansatza}
\eeq
and
\beq
%{\rm and}\quad\quad
R\sp{-\alpha\alpha}(u,v)=w_\alpha(v-u)
e\sp{-(\zeta(v-u)-\zeta(v)+\zeta(u))\alpha}.
\label{eq:ansatzb}
\eeq
For any Lie algebra, this ansatz solves (\ref{eq:spa})
and reduces (\ref{eq:spb}) to the equation
\beq
\label{eq:reduction}
\alpha\cdot {R\sp{\beta}(u,v)\over w_\beta(v)} -
\beta\cdot  {R\sp{\alpha}(v,u)\over w_\alpha(u)}
=c_{\alpha\beta}\sp\gamma .
\eeq
The consistency conditions exploited in the last section
are implicit here in the structure constants $c_{\alpha\beta}\sp\gamma$.
Certainly, for the case of $gl_n$ we may
solve (\ref{eq:reduction}) by setting
\beq
R\sp{i\alpha}(u,v)={1\over2}w_\alpha(v)\quad\quad
{\rm whenever}\quad \alpha\cdot e_i \not= 0
\label{eq:ansatzc}
\eeq
and zero otherwise. The remainder of this section will be devoted to
proving some of these assertions.

First let us show  that (\ref{eq:spa}) is satisfied. Now
\begin{displaymath}
R\sp{-\alpha\alpha}(u,v){w_\alpha(u)\over w_\alpha(v)}=
{{\sigma(u-\alpha)\sigma(v)\sigma(v-u-\alpha)}\over
{\sigma(u)\sigma(v-\alpha)\sigma(\alpha)\sigma(v-u)}}
=\zeta(-u)+\zeta(v-\alpha)+\zeta(\alpha)-\zeta(v-u).
\end{displaymath}
The exponential factors in our ansatz for
$R\sp{-\alpha\alpha}(u,v)$ have been chosen so that there
is cancellation leaving only the $\sigma$ factors in the
middle term. The final equality makes use of the
identity\cite{WW}
\begin{displaymath}
{\sigma(x+y)\sigma(y+z)\sigma(z+x)\over\sigma(x)\sigma(y)\sigma(z)\sigma(x+y+z)}
= \zeta(x)+\zeta(y)+\zeta(z)-\zeta(x+y+z).
\end{displaymath}
Finally, upon making use of
${w_\alpha\sp\prime(v) /w_\alpha(v)}=\zeta(v)-\zeta(\alpha)-\zeta(v-\alpha)$,
we  find (\ref{eq:spa}) holds for our choice of $\eta$.

{
As for (\ref{eq:spb}), first observe that
\begin{eqnarray*}
\lefteqn{
R\sp{-\alpha\alpha}(v,u)w_{\alpha+\beta}(v)+
R\sp{-\beta\beta}(u,v)w_{\alpha+\beta}(u) }\\
&=&
\Bigl[
-{{\sigma(u-v-\alpha)\sigma(v-\alpha-\beta)}\over {\sigma(\alpha)\sigma(v)}}
+
{{\sigma(v-u-\beta)\sigma(u-\alpha-\beta)}\over {\sigma(\beta)\sigma(u)}}
\Bigl]
{{e\sp{\zeta(u)\alpha+\zeta(v)\beta}}\over{\sigma(\alpha+\beta)\sigma(v-u)}}\cr
&=&
w_{\alpha}(u)w_{\beta}(v).
\end{eqnarray*}
\goodbreak}
\noindent
To obtain the final equality, we have employed the
\lq three-term equation\rq\ of Weierstrass\cite[\S$20\cdot53$]{WW},
\begin{eqnarray}
\sigma(x-y)\sigma(x+y)\sigma(z-t)\sigma(z+t)+
\sigma(y-z)\sigma(y+z)\sigma(x-t)\sigma(x+t)\cr
\nonumber
+\sigma(z-x)\sigma(z+x)\sigma(y-t)\sigma(y+t)=0.
\end{eqnarray}
This observation means that (\ref{eq:spb}) reduces to
(\ref{eq:reduction}).

\section{Discussion}

This paper has further investigated the $R$-matrix structure of the
Calogero-Moser models under the assumption of momentum independence.
We have shown that for the usual $L$-operator
($L=L\sp\dagger\Leftrightarrow w_\alpha=-w_{-\alpha}$)
and nondegenerate type IV potential,
no momentum independent $R$-matrix exists
whenever $n\ge4$.
Indeed our analysis
showed that  momentum independence actually gives
the $R$-matrices of \cite{AT} for the type I-III potentials
when the otherwise arbitrary projection operator $P\sp{ij}$
is chosen to vanish. For $n=2$ and $3$, solutions
may however be found for the type IV potential.

By enlarging the class of $L$-operators under consideration
to the family considered by Krichever, we were able to
construct an  appropriate  spectral parameter-dependent
$R$-matrix. This was given by
(\ref{eq:ansatza},\ref{eq:ansatzb},\ref{eq:ansatzc}).
We have worked throughout in terms of a basis of the Lie
algebra. This  has the merit of reducing  the
problem to  the two equations (\ref{eq:spa},\ref{eq:spb})
and highlighting the implicit consistency conditions.
The ansatz presented  by (\ref{eq:ansatza},\ref{eq:ansatzb}) is
independent of the Lie algebra to which $L$ is associated.
While we can certainly find consistent solutions to the
resulting  equation (\ref{eq:reduction}) in the case  of $gl_n$
we have not fully examined this  equation in the general
setting.

We must conclude by mentioning the very recent,
related work of Sklyanin\cite{Sklyanin93} which also constructs
$R$-matrices for the type IV Calogero-Moser model with spectral parameter.
At first glance our solutions are different and we have delayed
the written presentation of this work in order to clarify this
point. Certainly Sklyanin's approach is very different from our own.
We may easily verify that Sklyanin's ansatz satisfies our
(\ref{eq:ansatza},\ref{eq:ansatzb}) and so provides a solution.
The differences between the solutions has its origin
in our respective presentation
of the $L$-operators. Sklyanin in fact works with a conjugate
of Krichever's $L$-operator, $U L(u) U\sp{-1}$ where
$U_{ij}=e\sp{\zeta(u)x_i}\delta_{ij}$.
Once this is observed, our solutions are in fact in agreement,
our works providing independent proofs of this fact.

\vskip 0.5in
T.\,S. acknowledges financial support from both the
Daiwa Anglo-Japanese Foundation and Fuju-kai Foundation.
H.\,W.\,B. thanks A.\,J.\,Macfarlane for remarks about the importance
of spectral parameter dependent $R$-matrices.

%%%%%%%%%%%%%%%%%%%%%%%%%%%%%%%%%%%%%%%%%%%%%%%%%%%%%%%%%

\end{document}